# A Geometric Algebra Framework for a Multidimensional Analytic Signal

Dr. K. James Sangston
Sensors and Electromagnetic Applications Laboratory
Georgia Tech Research Institute
Georgia Institute of Technology
Atlanta, GA 30332-0858
404 407-7677
jim.sangston@gtri.gatech.edu; jsangston@mac.com

**Abstract**

This work examines the problem of extending the one-dimensional analytic signal, which is ubiquitous throughout signal processing, to higher dimensional signals. Bulow et al. and Felsberg et al. have previously used techniques from Clifford algebra and analysis to extend the one-dimensional analytic signal to higher dimensions. However, each author sets forth a different definition of a multidimensional analytic signal. Herein we follow an observation of Brackx et al. and adopt a general definition of an analytic signal that encompasses both the hypercomplex signal of Bulow et al. and the monogenic signal of Felsberg et al. within the same mathematical framework. The crux of our approach is captured by the following statement: A multidimensional analytic signal is generated by an idempotent. We develop this notion more specifically using examples from geometric algebra.

**Keywords**

multidimensional, analytic signal, Hilbert transform, idempotent, geometric algebra, orientation

**Introduction**

This work examines the problem of extending the one-dimensional analytic signal, which is ubiquitous throughout one-dimensional signal processing, to higher dimensional signals. It is motivated by the beautiful results of Bulow et al. [1] defining the hypercomplex signal and Felsberg et al. [3] defining the monogenic signal (and the respective dissertations underlying their works [2,4]). In these works the respective authors use techniques from Clifford algebra and analysis [5] to extend the one-dimensional analytic signal to higher dimensions. However, each author sets forth a different definition of a multidimensional analytic signal. Motivated by an observation of Brackx et al. [6], herein we adopt a general definition of an analytic signal to show how both the hypercomplex signal and the monogenic signal – as well as the single-orthant analytic signal of Hahn [7] – may be obtained within the same mathematical framework.

Before presenting the multidimensional case we motivate our approach by examining the one-dimensional problem. Let $f: \mathbb{R} \to \mathbb{R}$ be a given function. The Hilbert transform of $f(x)$ is defined as

$$H(f)(x) = \frac{1}{\pi} PV \int_{-\infty}^{\infty} \frac{f(t)}{x-t} dt \tag{1}$$

Here the notation $PV$ indicates that the singularity in the integrand at $x = t$ is approached symmetrically. We may define an operator as:

$$\mathcal{H}_1[f] = iH(f) \tag{2}$$

It is a property of the Hilbert transform that

$$H(H(f)) = -f \tag{3}$$

Thus it follows that

$$\mathcal{H}_1[\mathcal{H}_1[f]] = iH(iH(f)) = -i^2 f = f \tag{4}$$

Therefore $\mathcal{H}_1$ is invertible with

$$\mathcal{H}_1^{-1} = \mathcal{H}_1 \tag{5}$$

We now define a function $f_{\mathcal{H}}: \mathbb{R} \to \mathbb{C}$ as

$$f_{\mathcal{H}_1} = \mathcal{H}_1[f] \tag{6}$$

Because of the invertibility of $\mathcal{H}_1$ we also have

$$f = \mathcal{H}_1[f_{\mathcal{H}_1}] \tag{7}$$

Thus the operator $\mathcal{H}_1$ may be thought of as toggling between $f$ and $f_{\mathcal{H}_1}$. The well-known one-dimensional analytic signal $f_A: \mathbb{R} \to \mathbb{C}$ is obtained by combining these signals:[1]

$$f_A(x) = \frac{1}{2} f(x) + \frac{1}{2} f_{\mathcal{H}_1}(x) \tag{8}$$

---

[1] The factor of $1/2$ is not generally included in engineering applications. When comparing the results herein with results in the engineering literature the need to account for this factor is assumed without further comment.

It follows immediately from the invertibility of $\mathcal{H}_1$ that we have

$$f_A = \mathcal{H}_1[f_A] \qquad (9)$$

Thus one may write

$$f_A = \frac{1}{2}(I + \mathcal{H}_1)f = \frac{1}{2}(I + \mathcal{H}_1)f_{\mathcal{H}_1} \qquad (10)$$

It appears that a useful aspect of the analytic signal $f_A$ is that it can be defined from either $f$ or $f_{\mathcal{H}_1}$. In this regard $f$ and $f_{\mathcal{H}_1}$ carry redundant information, and adding $f_{\mathcal{H}_1}$ does not add extraneous information. Moreover, knowledge of one is equivalent to knowledge of the other. Forming $f_A$ generally yields a convenient signal for defining such concepts as instantaneous frequency, amplitude, and phase of $f$.

The Fourier multiplier of the Hilbert transform is $-i\,sgn(\omega)$. Thus the Fourier multiplier of the operator $\mathcal{H}_1$ is

$$a_{\mathcal{H}_1}(\omega) = sgn(\omega) = \frac{\omega}{|\omega|} \qquad (11)$$

It satisfies

$$a_{\mathcal{H}_1}^2(\omega) = 1 \qquad (12)$$

Define

$$\psi_{\mathcal{H}_1}(\omega) = \frac{1 + a_{\mathcal{H}_1}(\omega)}{2} = \begin{cases} 1, \omega \geq 0 \\ 0, \omega < 0 \end{cases} \qquad (13)$$

We find in the Fourier domain

$$\mathcal{F}[f_A] = \frac{1}{2}\mathcal{F}[f] + \frac{1}{2}\mathcal{F}(\mathcal{H}_1[f]) = \psi_{\mathcal{H}_1}(\omega)\mathcal{F}[f](\omega) \qquad (14)$$

Based on this formulation an analytic signal in one dimension is sometimes defined as a signal whose negative frequencies vanish. However, for our purposes, the crucial observation is that the function $\psi_{\mathcal{H}_1}(\omega)$ is an idempotent, i.e. it squares to itself:

$$\psi_{\mathcal{H}_1}^2(\omega) = \psi_{\mathcal{H}_1}(\omega) \qquad (15)$$

The crux of our approach is thus captured by the following statement adapted from [6]:

*A multidimensional analytic signal is generated by an idempotent.*

Accordingly we define an analytic signal in higher dimensions in the frequency domain as

$$\mathcal{F}[f_A](\boldsymbol{\omega}) = \psi_{\mathcal{H}}(\boldsymbol{\omega})\mathcal{F}[f](\boldsymbol{\omega}) \qquad (16)$$

where $\psi_{\mathcal{H}}(\boldsymbol{\omega})$ is an idempotent function in an appropriate geometric algebra (to be defined below). Defining a multidimensional analytic signal in terms of idempotents may be viewed as the analogue of defining the one-dimensional analytic signal in terms of vanishing negative frequencies [6]. In what follows we develop this notion more specifically using examples from geometric algebra. We proceed formally through examples rather than mathematical rigor, which admittedly is absent.

**Geometric Algebra Framework**

Let $x = (x_1, \ldots, x_N)$ be an $N$-tuple of real numbers and let $f: \mathbb{R}^N \to \mathbb{R}$ be a real-valued function $f(x)$. We want to associate a multidimensional analytic signal with the real-valued $N$-dimensional signal $f$. To do so we first identify an appropriate geometric algebra. Let $L = 2m + 1 > N$ be an odd integer such that

$$\frac{L(L-1)}{2} = m(2m+1) \tag{17}$$

is odd. It follows that $m$ must be odd. Writing $m = 2n + 1$ we therefore have for $n = 0, 1, 2, \cdots$

$$L = 4n + 3 = 3, 7, 11, 15, \cdots \tag{18}$$

Let $\mathbb{G}_L$ be the real geometric algebra associated with $\mathbb{R}^L \supset \mathbb{R}^N$. It has orthogonal basis vectors $e_1, \ldots, e_L$ with the following properties:

$$e_k^2 = 1, k = 1, \ldots, L \tag{19}$$

$$e_k e_j = -e_j e_k, \; j, k = 1, \ldots, L, k \neq j \tag{20}$$

Define an ordered multi-index $\mathcal{A}_k = \{i_1, \ldots, i_k\}$ where the indices $i_1, \ldots, i_k$ are distinct integers chosen $k$ at a time from the set $\{1, \ldots, L\}$. A general multivector $M \in \mathbb{G}_L$ may be written as

$$M = \gamma_0 + \sum_{k=1}^{L} \sum_{|\mathcal{A}_k|} \gamma_{\mathcal{A}_k} e_{\mathcal{A}_k} \tag{21}$$

Here $\gamma_0, \gamma_{\mathcal{A}_k}$ are real numbers and

$$e_{\mathcal{A}_k} = e_{i_1} \ldots e_{i_k} \tag{22}$$

A sum over $|\mathcal{A}_k|$ has $\binom{L}{k}$ terms. We denote the unit pseudoscalar of $\mathbb{G}_L$ as

$$I_L = e_1 \ldots e_L \tag{23}$$

For $L$ as given above the unit pseudoscalar has the properties that it commutes with all elements of $\mathbb{G}_L$ and

$$I_L^2 = -1 \tag{24}$$

We write for a real vector $y \in \mathbb{R}^L \subset \mathbb{G}_L$

$$y = \sum_{k=1}^{L} y_k e_k \tag{25}$$

The square of a real vector in $\mathbb{G}_L$ is given by

$$y^2 = |y|^2 = \sum_{i=1}^{L} y_k^2 > 0 \tag{26}$$

The scalar product of two real vectors is given by

$$\langle y, v \rangle = y \cdot v = \sum_{k=1}^{L} y_k v_k \tag{27}$$

Here $y \cdot v$ is the standard Euclidean inner product. Finally, for a function $g: \mathbb{R}^N \to \mathbb{G}_L$ we use the multidimensional Fourier transform[2] as follows:

$$\mathcal{F}[g](\boldsymbol{\omega}) = \int_{\mathbb{R}^N} g(\boldsymbol{x}) e^{-I_L \boldsymbol{x} \cdot \boldsymbol{\omega}} d\boldsymbol{x} \tag{28}$$

Here and in the following $\boldsymbol{x}, \boldsymbol{\omega} \in \mathbb{R}^N \subset \mathbb{R}^L$ are real vectors. This has the form of the ordinary Fourier transform with the unit imaginary $i$ replaced by the unit pseudoscalar $I_L$.

Let $\psi_{\mathcal{H}}: \mathbb{R}^N \to \mathbb{G}_L$ be an idempotent of $\mathbb{G}_L$ and define a Fourier multiplier $a_{\mathcal{H}}: \mathbb{R}^N \to \mathbb{G}_L$ as

$$a_{\mathcal{H}}(\boldsymbol{\omega}) = 2\psi_{\mathcal{H}}(\boldsymbol{\omega}) - 1 \tag{29}$$

Because $\psi_{\mathcal{H}}(\boldsymbol{\omega})$ is an idempotent, it follows that

$$a_{\mathcal{H}}^2(\boldsymbol{\omega}) = (2\psi_{\mathcal{H}}(\boldsymbol{\omega}) - 1)^2 = 4\psi_{\mathcal{H}}^2(\boldsymbol{\omega}) - 4\psi_{\mathcal{H}}(\boldsymbol{\omega}) + 1 = 1 \tag{30}$$

To specify an idempotent in $\mathbb{G}_L$ we may specify the corresponding Fourier multiplier $a_{\mathcal{H}}(\boldsymbol{\omega})$.

Let $A$ be a specific set of frequency values $\mathbb{R}^N, N \geq 1$ and $A^C$ be the complement of $A$. We may define an appropriate Fourier multiplier as a scalar-valued function

$$a_{\mathcal{H}}(\boldsymbol{\omega}) = \begin{cases} 1, \boldsymbol{\omega} \in A \\ -1, \boldsymbol{\omega} \in A^C \end{cases} \tag{31}$$

With this choice of multiplier the scalar-valued idempotent in $\mathbb{G}_L$ then becomes

$$\psi_{\mathcal{H}}(\boldsymbol{\omega}) = \begin{cases} 1, \boldsymbol{\omega} \in A \\ 0, \boldsymbol{\omega} \in A^C \end{cases} \tag{32}$$

Alternatively let $a_{\mathcal{H}}(\boldsymbol{\omega})$ be a multivector in $\mathbb{G}_L$ having no scalar component and denote the odd-grade and even-grade components of $a_{\mathcal{H}}(\boldsymbol{\omega})$ as $a_{\mathcal{H},-}(\boldsymbol{\omega})$ and $a_{\mathcal{H},+}(\boldsymbol{\omega})$ respectively:

$$a_{\mathcal{H}}(\boldsymbol{\omega}) = a_{\mathcal{H},-}(\boldsymbol{\omega}) + a_{\mathcal{H},+}(\boldsymbol{\omega}) \tag{33}$$

We now assume the following conditions are satisfied: [3]

$$a_{\mathcal{H},-}(\boldsymbol{\omega}) = \|a_{\mathcal{H},-}(\boldsymbol{\omega})\| \hat{a}_{\mathcal{H},-}(\boldsymbol{\omega}) \tag{34}$$

$$a_{\mathcal{H},+}(\boldsymbol{\omega}) = \|a_{\mathcal{H},+}(\boldsymbol{\omega})\| \hat{a}_{\mathcal{H},+}(\boldsymbol{\omega}) \tag{35}$$

where $\|a_{\mathcal{H},-}(\boldsymbol{\omega})\| \geq 0, \|a_{\mathcal{H},+}(\boldsymbol{\omega})\| \geq 0$ are real numbers satisfying

$$\|a_{\mathcal{H},-}(\boldsymbol{\omega})\|^2 - \|a_{\mathcal{H},+}(\boldsymbol{\omega})\|^2 = 1 \tag{36}$$

---

[2] It is possible to define the Fourier transform in other ways in the context of Clifford algebra. We leave investigation of the consequences of other definitions for further investigation.
[3] These are sufficient conditions to insure that the related function $\psi_{\mathcal{H}}(\boldsymbol{\omega})$ is an idempotent in $\mathbb{G}_L$; we do not know if they are necessary.

Here $\hat{a}_{\mathcal{H},-}(\boldsymbol{\omega}), \hat{a}_{\mathcal{H},+}(\boldsymbol{\omega})$ are unit multivectors satisfying

$$\hat{a}_{\mathcal{H},-}^2(\boldsymbol{\omega}) = 1 \tag{37}$$

$$\hat{a}_{\mathcal{H},+}^2(\boldsymbol{\omega}) = -1 \tag{38}$$

$$\hat{a}_{\mathcal{H},-}(\boldsymbol{\omega})\hat{a}_{\mathcal{H},+}(\boldsymbol{\omega}) + \hat{a}_{\mathcal{H},+}(\boldsymbol{\omega})\hat{a}_{\mathcal{H},-}(\boldsymbol{\omega}) = 0 \tag{39}$$

With these conditions we find

$$a_{\mathcal{H}}^2(\boldsymbol{\omega})$$
$$= \|a_{\mathcal{H},-}(\boldsymbol{\omega})\|^2 + \|a_{\mathcal{H},-}(\boldsymbol{\omega})\|\|a_{\mathcal{H},+}(\boldsymbol{\omega})\|\left(\hat{a}_{\mathcal{H},-}(\boldsymbol{\omega})\hat{a}_{\mathcal{H},+}(\boldsymbol{\omega}) + \hat{a}_{\mathcal{H},+}(\boldsymbol{\omega})\hat{a}_{\mathcal{H},-}(\boldsymbol{\omega})\right)$$
$$- \|a_{\mathcal{H},+}(\boldsymbol{\omega})\|^2$$
$$= 1 \tag{40}$$

Because the odd-grade component of the multiplier must have magnitude greater than or equal to 1, the multiplier is a unit multivector if and only if the even-grade component is identically equal to 0. As a result we may write the even grade component as

$$a_{\mathcal{H},+}(\boldsymbol{\omega}) = s(\boldsymbol{\omega})\sqrt{\|a_{\mathcal{H},-}(\boldsymbol{\omega})\|^2 - 1}\,\hat{a}_{\mathcal{H},+}(\boldsymbol{\omega}) \tag{41}$$

Here $s(\boldsymbol{\omega})$ is the sign of the square root, which may be chosen freely. The idempotent is now given by

$$\psi_{\mathcal{H}}(\boldsymbol{\omega}) = \frac{1 + a_{\mathcal{H}}(\boldsymbol{\omega})}{2}$$
$$= \frac{1 + \|a_{\mathcal{H},-}(\boldsymbol{\omega})\|\hat{a}_{\mathcal{H},-}(\boldsymbol{\omega}) + s(\boldsymbol{\omega})\sqrt{\|a_{\mathcal{H},-}(\boldsymbol{\omega})\|^2 - 1}\,\hat{a}_{\mathcal{H},+}(\boldsymbol{\omega})}{2} \tag{42}$$

For concreteness in what follows, we now specify a class of idempotents in $\mathbb{G}_L$ that we use throughout the remainder of the paper. First, let $m(\boldsymbol{\omega})$ be a scalar-valued function taking values

$$a_{\mathcal{H}}(\boldsymbol{\omega}) = m(\boldsymbol{\omega}) = \pm 1 \tag{43}$$

The idempotent becomes

$$\psi_{\mathcal{H}}(\boldsymbol{\omega}) = \frac{1 + m(\boldsymbol{\omega})}{2} \tag{44}$$

Alternatively let $\boldsymbol{v} \in \mathbb{R}^N$ be a vector and $I_{L-1} \in \mathbb{G}_L$ be a pseudovector given respectively by

$$\boldsymbol{v}(\boldsymbol{\omega}) = \sum_{i=1}^N v_i(\boldsymbol{\omega})\,\boldsymbol{e}_i \tag{45}$$

$$I_{L-1} = I_L \boldsymbol{e}_L = \boldsymbol{e}_1 \ldots \boldsymbol{e}_{L-1} \tag{46}$$

where $\|\boldsymbol{v}\| \geq 1$. Note that $I_{L-1}$ has even grade and includes all basis vectors $\boldsymbol{e}_1, \ldots, \boldsymbol{e}_N$ of $\mathbb{R}^N$. It satisfies

$$I_{L-1}^2 = I_L^2 \boldsymbol{e}_L^2 = -1 \tag{47}$$

$$\boldsymbol{v} \wedge I_{L-1} = 0 \tag{48}$$

We now set

$$a_{\mathcal{H},-}(\boldsymbol{\omega}) = \boldsymbol{v}(\boldsymbol{\omega}) \tag{49}$$

$$a_{\mathcal{H},+}(\boldsymbol{\omega}) = P(\boldsymbol{\omega}) I_{L-1} \tag{50}$$

where

$$P(\boldsymbol{\omega}) = s(\boldsymbol{\omega})\sqrt{\|\boldsymbol{v}(\boldsymbol{\omega})\|^2 - 1} \tag{51}$$

The Fourier multiplier becomes

$$a_{\mathcal{H}}(\boldsymbol{\omega}) = \boldsymbol{v}(\boldsymbol{\omega}) + P(\boldsymbol{\omega}) I_{L-1} \tag{52}$$

The idempotent becomes

$$\psi_{\mathcal{H}}(\boldsymbol{\omega}) = \tfrac{1+\boldsymbol{v}(\boldsymbol{\omega})+P(\boldsymbol{\omega}) I_{L-1}}{2} \tag{53}$$

**A Hierarchy of Multidimensional Analytic Signals**

We now use the idempotents specified above to develop a hierarchy of multidimensional analytic signals associated with a given real-valued function $f(\boldsymbol{x}), \boldsymbol{x} \in \mathbb{R}^N$. We begin by defining the components of $f$ having even and odd symmetry with respect to the argument $\boldsymbol{x}$ as

$$f_e(\boldsymbol{x}) = \tfrac{f(\boldsymbol{x})+f(-\boldsymbol{x})}{2} \tag{54}$$

$$f_o(\boldsymbol{x}) = \tfrac{f(\boldsymbol{x})-f(-\boldsymbol{x})}{2} \tag{55}$$

We may thus write

$$f(\boldsymbol{x}) = f_e(\boldsymbol{x}) + f_o(\boldsymbol{x}) \tag{56}$$

The Fourier transform becomes

$$\mathcal{F}[f](\boldsymbol{\omega}) = \int_{\mathbb{R}^N} f_e(\boldsymbol{x}) \cos(\boldsymbol{x} \cdot \boldsymbol{\omega}) \, d\boldsymbol{x} - I_L \int_{\mathbb{R}^N} f_o(\boldsymbol{x}) \sin(\boldsymbol{x} \cdot \boldsymbol{\omega}) \, d\boldsymbol{x}$$

$$= F_e(\boldsymbol{\omega}) - I_L F_o(\boldsymbol{\omega}) \tag{57}$$

The scalar part of the Fourier transform has even symmetry with respect to the argument whereas the pseudoscalar part has odd symmetry. Depending on how the idempotent is defined, in the frequency domain we now find alternatively for the product of the multiplier $a_{\mathcal{H}}$ with the Fourier transform $\mathcal{F}[f]$:

$$a_{\mathcal{H}}(\boldsymbol{\omega})\mathcal{F}[f](\boldsymbol{\omega}) = \begin{cases} m(\boldsymbol{\omega})F_e(\boldsymbol{\omega}) - I_L m(\boldsymbol{\omega})F_o(\boldsymbol{\omega}) \\ \boldsymbol{v}(\boldsymbol{\omega})F_e(\boldsymbol{\omega}) + P(\boldsymbol{\omega})F_e(\boldsymbol{\omega}) I_{L-1} - I_L \boldsymbol{v}(\boldsymbol{\omega})F_o(\boldsymbol{\omega}) + P(\boldsymbol{\omega})F_o(\boldsymbol{\omega})\boldsymbol{e}_L \end{cases} \tag{58}$$

Here we have used the fact

$$I_L I_{L-1} = -\boldsymbol{e}_L \tag{59}$$

Define the components of $m(\boldsymbol{\omega}), v(\boldsymbol{\omega})$ and $P(\boldsymbol{\omega})$ having even and odd symmetry with respect to the argument $\boldsymbol{\omega}$ simlarly as in (54-55). Also define

$$F_{\mathcal{H},-,e}(\boldsymbol{\omega}) = v_e(\boldsymbol{\omega})F_e(\boldsymbol{\omega}) + P_o(\boldsymbol{\omega})F_o(\boldsymbol{\omega})e_L \tag{60}$$

$$F_{\mathcal{H},+,e}(\boldsymbol{\omega}) = P_e(\boldsymbol{\omega})F_e(\boldsymbol{\omega})\, I_{L-1} - I_L v_o(\boldsymbol{\omega})F_o(\boldsymbol{\omega}) \tag{61}$$

$$F_{\mathcal{H},-,o}(\boldsymbol{\omega}) = v_o(\boldsymbol{\omega})F_e(\boldsymbol{\omega}) + P_e(\boldsymbol{\omega})F_o(\boldsymbol{\omega})e_L \tag{62}$$

$$F_{\mathcal{H},+,o}(\boldsymbol{\omega}) = P_o(\boldsymbol{\omega})F_e(\boldsymbol{\omega})\, I_{L-1} - I_L v_e(\boldsymbol{\omega})F_o(\boldsymbol{\omega}) \tag{63}$$

The subscript indicates even (+) or odd (−) grade as well as even (e) or odd (o) symmetry with respect to the argument. We now find

$$a_{\mathcal{H}}(\boldsymbol{\omega})\mathcal{F}[f_{\mathcal{H}}](\boldsymbol{\omega}) = \begin{cases} m_e(\boldsymbol{\omega})F_e(\boldsymbol{\omega}) - I_L m_o(\boldsymbol{\omega})F_o(\boldsymbol{\omega}) + m_o(\boldsymbol{\omega})F_e(\boldsymbol{\omega}) - I_L m_e(\boldsymbol{\omega})F_o(\boldsymbol{\omega}) \\ F_{\mathcal{H},-,e}(\boldsymbol{\omega}) + F_{\mathcal{H},+,e}(\boldsymbol{\omega}) + F_{\mathcal{H},-,o}(\boldsymbol{\omega}) + F_{\mathcal{H},+,o}(\boldsymbol{\omega}) \end{cases} \tag{64}$$

*Generalized Analytic Signal*

In the frequency domain we define a <u>generalized analytic signal</u> $f_A$ associated with a function $f$ as

$$\mathcal{F}[f_A](\boldsymbol{\omega}) = \psi_{\mathcal{H}}(\boldsymbol{\omega})\mathcal{F}[f](\boldsymbol{\omega}) = \tfrac{1}{2}\mathcal{F}[f](\boldsymbol{\omega}) + \tfrac{a_{\mathcal{H}}(\boldsymbol{\omega})}{2}\mathcal{F}[f](\boldsymbol{\omega}) \tag{65}$$

Assume[4] a suitable operator $\mathcal{H}$ exists such that upon taking the inverse Fourier transform we have

$$\mathcal{H}[f](\boldsymbol{x}) = \mathcal{F}^{-1}[a_{\mathcal{H}}(\boldsymbol{\omega})\mathcal{F}[f](\boldsymbol{\omega})](\boldsymbol{x}) \tag{66}$$

Because $a_{\mathcal{H}}^2(\boldsymbol{\omega}) = 1$, applying the operator twice in the frequency domain is equivalent to the identity:

$$\mathcal{H}^{-1} = \mathcal{H} \tag{67}$$

We call such an operator $\mathcal{H}$ an <u>extended Hilbert operator</u>. We now find

$$f_A(\boldsymbol{x}) = \tfrac{1}{2}f(\boldsymbol{x}) + \tfrac{1}{2}\mathcal{H}[f](\boldsymbol{x}) \tag{68}$$

In general $f_A(\boldsymbol{x})$ is a function from $\mathbb{R}^N$ to $\mathbb{G}_L$. Define a function $f_{\mathcal{H}}: \mathbb{R}^N \to \mathbb{G}_L$ by

$$f_{\mathcal{H}} = \mathcal{H}[f] \tag{69}$$

We call $f_{\mathcal{H}}$ the <u>extended Hilbert transform</u>. Because of the invertibility of the operator $\mathcal{H}$, we have

$$f = \mathcal{H}[f_{\mathcal{H}}] \tag{70}$$

Application of $\mathcal{H}$ toggles between $f$ and $f_{\mathcal{H}}$ and thus mimics the corresponding property in the one-dimensional case. The generalized analytic function $f_A$ is now given by

$$f_A(\boldsymbol{x}) = \tfrac{1}{2}f(\boldsymbol{x}) + \tfrac{1}{2}f_{\mathcal{H}}(\boldsymbol{x}) \tag{71}$$

---

[4] An appropriate (convolutional) operator need not exist for an arbitrary idempotent. Below we give examples, but we leave further exploration of this issue to another time.

It is straightforward to see that $f_A$ satisfies:

$$f_A = \mathcal{H}[f_A] \tag{72}$$

This definition of a generalized analytic signal extends the one-dimensional analytic signal to multiple dimensions via the mechanism of the idempotent.

For the multipliers (equivalently idempotents) given above the extended Hilbert transform may be written alternatively as

$$f_\mathcal{H}(\boldsymbol{x}) =$$
$$\mathcal{F}^{-1}[m_e(\boldsymbol{\omega})F_e(\boldsymbol{\omega})](\boldsymbol{x}) - I_L\mathcal{F}^{-1}[m_o(\boldsymbol{\omega})F_o(\boldsymbol{\omega})](\boldsymbol{x}) + \mathcal{F}^{-1}[m_o(\boldsymbol{\omega})F_e(\boldsymbol{\omega})](\boldsymbol{x}) - I_L\mathcal{F}^{-1}[m_e(\boldsymbol{\omega})F_o(\boldsymbol{\omega})](\boldsymbol{x}) \tag{73}$$

or as

$$f_\mathcal{H}(\boldsymbol{x}) = \mathcal{F}^{-1}[F_{\mathcal{H},-,e}(\boldsymbol{\omega})](\boldsymbol{x}) + \mathcal{F}^{-1}[F_{\mathcal{H},+,e}(\boldsymbol{\omega})](\boldsymbol{x}) + \mathcal{F}^{-1}[F_{\mathcal{H},-,o}(\boldsymbol{\omega})](\boldsymbol{x}) + \mathcal{F}^{-1}[F_{\mathcal{H},+,o}(\boldsymbol{\omega})](\boldsymbol{x}) \tag{74}$$

Examine each term in $f_\mathcal{H}(\boldsymbol{x})$ separately.

For the scalar multiplier we find (various terms vanish due to a mismatch of symmetries with respect to $\boldsymbol{\omega}$)

$$\mathcal{F}^{-1}[m_e(\boldsymbol{\omega})F_e(\boldsymbol{\omega})](\boldsymbol{x}) = \int_{\mathbb{R}^N} m_e(\boldsymbol{\omega})F_e(\boldsymbol{\omega})\cos(\boldsymbol{x}\cdot\boldsymbol{\omega})\,d\boldsymbol{\omega} \tag{75}$$

$$\mathcal{F}^{-1}[m_o(\boldsymbol{\omega})F_o(\boldsymbol{\omega})](\boldsymbol{x}) = -I_L \int_{\mathbb{R}^N} m_o(\boldsymbol{\omega})F_o(\boldsymbol{\omega})\cos(\boldsymbol{x}\cdot\boldsymbol{\omega})\,d\boldsymbol{\omega} \tag{76}$$

$$\mathcal{F}^{-1}[m_o(\boldsymbol{\omega})F_e(\boldsymbol{\omega})](\boldsymbol{x}) = I_L \int_{\mathbb{R}^N} m_o(\boldsymbol{\omega})F_e(\boldsymbol{\omega})\sin(\boldsymbol{x}\cdot\boldsymbol{\omega})\,d\boldsymbol{\omega} \tag{77}$$

$$\mathcal{F}^{-1}[m_e(\boldsymbol{\omega})F_o(\boldsymbol{\omega})](\boldsymbol{x}) = \int_{\mathbb{R}^N} m_e(\boldsymbol{\omega})F_o(\boldsymbol{\omega})\sin(\boldsymbol{x}\cdot\boldsymbol{\omega})\,d\boldsymbol{\omega} \tag{78}$$

In this case the extended Hilbert transform becomes

$$f_\mathcal{H}(\boldsymbol{x}) = f_{\mathcal{H},Re}(\boldsymbol{x}) + I_L f_{\mathcal{H},Im}(\boldsymbol{x}) \tag{79}$$

where

$$f_{\mathcal{H},Re}(\boldsymbol{x}) = \int_{\mathbb{R}^N} m_e(\boldsymbol{\omega})\{F_e(\boldsymbol{\omega})\cos(\boldsymbol{x}\cdot\boldsymbol{\omega}) + F_o(\boldsymbol{\omega})\sin(\boldsymbol{x}\cdot\boldsymbol{\omega})\}\,d\boldsymbol{\omega} \tag{80}$$

$$f_{\mathcal{H},Im}(\boldsymbol{x}) = \int_{\mathbb{R}^N} m_o(\boldsymbol{\omega})\{F_e(\boldsymbol{\omega})\sin(\boldsymbol{x}\cdot\boldsymbol{\omega}) - F_o(\boldsymbol{\omega})\cos(\boldsymbol{x}\cdot\boldsymbol{\omega})\}\,d\boldsymbol{\omega} \tag{81}$$

Alternatively, for the multiplier formed from vector and pseudovector components we find

$$\mathcal{F}^{-1}[F_{\mathcal{H},-,e}(\boldsymbol{\omega})](\boldsymbol{x}) = \int_{\mathbb{R}^N} \boldsymbol{v}_e(\boldsymbol{\omega})F_e(\boldsymbol{\omega})\cos(\boldsymbol{x}\cdot\boldsymbol{\omega})\,d\boldsymbol{\omega} + \int_{\mathbb{R}^N} P_o(\boldsymbol{\omega})F_o(\boldsymbol{\omega})\cos(\boldsymbol{x}\cdot\boldsymbol{\omega})\,d\boldsymbol{\omega}\,\boldsymbol{e}_L \quad (82)$$

$$\mathcal{F}^{-1}[F_{\mathcal{H},+,e}(\boldsymbol{\omega})](\boldsymbol{x}) = I_L \int_{\mathbb{R}^N} P_e(\boldsymbol{\omega})F_e(\boldsymbol{\omega})\cos(\boldsymbol{x}\cdot\boldsymbol{\omega})\,d\boldsymbol{\omega}\,\boldsymbol{e}_L - I_L \int_{\mathbb{R}^N} \boldsymbol{v}_o(\boldsymbol{\omega})F_o(\boldsymbol{\omega})\cos(\boldsymbol{x}\cdot\boldsymbol{\omega})\,d\boldsymbol{\omega} \quad (83)$$

$$\mathcal{F}^{-1}[F_{\mathcal{H},-,o}(\boldsymbol{\omega})](\boldsymbol{x}) = I_L \int_{\mathbb{R}^N} \boldsymbol{v}_o(\boldsymbol{\omega})F_e(\boldsymbol{\omega})\sin(\boldsymbol{x}\cdot\boldsymbol{\omega})\,d\boldsymbol{\omega} + I_L \int_{\mathbb{R}^N} P_e(\boldsymbol{\omega})F_o(\boldsymbol{\omega})\sin(\boldsymbol{x}\cdot\boldsymbol{\omega})\,d\boldsymbol{\omega}\,\boldsymbol{e}_L \quad (84)$$

$$\mathcal{F}^{-1}[F_{\mathcal{H},+,o}(\boldsymbol{\omega})](\boldsymbol{x}) = -\int_{\mathbb{R}^N} P_o(\boldsymbol{\omega})F_e(\boldsymbol{\omega})\sin(\boldsymbol{x}\cdot\boldsymbol{\omega})\,d\boldsymbol{\omega}\,\boldsymbol{e}_L + \int_{\mathbb{R}^N} \boldsymbol{v}_e(\boldsymbol{\omega})F_o(\boldsymbol{\omega})\sin(\boldsymbol{x}\cdot\boldsymbol{\omega})\,d\boldsymbol{\omega} \quad (85)$$

The extended Hilbert transform becomes

$$f_{\mathcal{H}}(\boldsymbol{x}) = I_L \mathbb{v}(\boldsymbol{x}) + \mathbb{u}(\boldsymbol{x}) \quad (86)$$

where

$$\mathbb{v}(\boldsymbol{x}) = \sum_{i=1}^{N} f_{\mathcal{H},i}(\boldsymbol{x})\,\boldsymbol{e}_i + f_{\mathcal{H},N+1}(\boldsymbol{x})\,\boldsymbol{e}_L \quad (87)$$

$$\mathbb{u}(\boldsymbol{x}) = \sum_{i=1}^{N} f_{\mathcal{H},N+1+i}(\boldsymbol{x})\,\boldsymbol{e}_i + f_{\mathcal{H},2(N+1)}(\boldsymbol{x})\,\boldsymbol{e}_L \quad (88)$$

with

$$f_{\mathcal{H},i}(\boldsymbol{x}) = \int_{\mathbb{R}^N} v_{i,o}(\boldsymbol{\omega})(F_e(\boldsymbol{\omega})\sin(\boldsymbol{x}\cdot\boldsymbol{\omega}) - F_o(\boldsymbol{\omega})\cos(\boldsymbol{x}\cdot\boldsymbol{\omega}))d\boldsymbol{\omega}\,, i=1,\ldots,N \quad (89)$$

$$f_{\mathcal{H},N+1}(\boldsymbol{x}) = \int_{\mathbb{R}^N} P_e(\boldsymbol{\omega})(F_e(\boldsymbol{\omega})\cos(\boldsymbol{x}\cdot\boldsymbol{\omega}) + F_o(\boldsymbol{\omega})\sin(\boldsymbol{x}\cdot\boldsymbol{\omega}))d\boldsymbol{\omega} \quad (90)$$

$$f_{\mathcal{H},N+1+i}(\boldsymbol{x}) = \int_{\mathbb{R}^N} v_{i,e}(\boldsymbol{\omega})\{F_e(\boldsymbol{\omega})\cos(\boldsymbol{x}\cdot\boldsymbol{\omega}) + F_o(\boldsymbol{\omega})\sin(\boldsymbol{x}\cdot\boldsymbol{\omega})\}d\boldsymbol{\omega}\,, i=1,\ldots,N \quad (91)$$

$$f_{\mathcal{H},2(N+1)}(\boldsymbol{x}) = \int_{\mathbb{R}^N} P_o(\boldsymbol{\omega})\{F_o(\boldsymbol{\omega})\cos(\boldsymbol{x}\cdot\boldsymbol{\omega}) - F_e(\boldsymbol{\omega})\sin(\boldsymbol{x}\cdot\boldsymbol{\omega})\}d\boldsymbol{\omega} \quad (92)$$

For an idempotent generated by a scalar function, $f_{\mathcal{H}}(\boldsymbol{x})$ comprises a scalar and a pseudoscalar term, whereas for an idempotent generated by vector and pseudovector functions, $f_{\mathcal{H}}(\boldsymbol{x})$ comprises a vector term and a pseudovector term. The generalized analytic signal may now be written as

$$f_A(\boldsymbol{x}) = \begin{cases} \frac{1}{2}f(\boldsymbol{x}) + \frac{1}{2}I_L f_{\mathcal{H},Im}(\boldsymbol{x}) + \frac{1}{2}f_{\mathcal{H},Re}(\boldsymbol{x}) \\ \frac{1}{2}f(\boldsymbol{x}) + \frac{1}{2}I_L \mathbb{v}(\boldsymbol{x}) + \frac{1}{2}\mathbb{u}(\boldsymbol{x}) \end{cases} \quad (93)$$

The extended Hilbert transform is completely determined by the scalar fields $f_{\mathcal{H},Re}(\boldsymbol{x}), f_{\mathcal{H},Im}(\boldsymbol{x})$ or alternatively by the vector fields $\mathbb{u}(\boldsymbol{x}), \mathbb{v}(\boldsymbol{x})$. Moreover, as discussed above the original function $f(\boldsymbol{x})$ can be obtained from the extended Hilbert transform by applying the operator $\mathcal{H}$. Consequently in some applications it may be beneficial to use either $f_{\mathcal{H},Re}(\boldsymbol{x}), f_{\mathcal{H},Im}(\boldsymbol{x})$ or $\mathbb{u}(\boldsymbol{x}), \mathbb{v}(\boldsymbol{x})$ in place of the original function $f(\boldsymbol{x})$.

*Generic Analytic Signal*

To restrict the generalized analytic signal to a smaller subset of signals we now impose the following symmetry conditions on the Fourier multiplier $a_\mathcal{H}(\boldsymbol{\omega})$:

$$m_e(\boldsymbol{\omega}) \equiv 0 \tag{94}$$

$$P_o(\boldsymbol{\omega}) \equiv 0 \tag{95}$$

$$\boldsymbol{v}_e(\boldsymbol{\omega}) \equiv 0 \tag{96}$$

In the scalar formulation the multiplier is odd whereas in the vector-pseudovector formulation the odd-grade component (in this case a vector) of the multiplier is odd and the even-grade component (in this case a pseudovector) is even. These conditions imply

$$f_{\mathcal{H},Re}(\boldsymbol{x}) = 0 \tag{97}$$

$$\mathbb{u}(\boldsymbol{x}) = 0 \tag{98}$$

The extended Hilbert transform becomes

$$f_\mathcal{H}(\boldsymbol{x}) = \begin{cases} I_L sgn\left(f_{\mathcal{H},Im}(\boldsymbol{x})\right)|f_{\mathcal{H},Im}(\boldsymbol{x})| \\ I_L \hat{\mathbb{v}}(\boldsymbol{x}) \|\mathbb{v}(\boldsymbol{x})\| \end{cases} \tag{99}$$

where

$$\hat{\mathbb{v}}(\boldsymbol{x}) = \sum_{i=1}^{N} \frac{f_{\mathcal{H},i}(\boldsymbol{x})}{\sqrt{\sum_{i=1}^{N+1}|f_{\mathcal{H},i}(\boldsymbol{x})|^2}} \boldsymbol{e}_i + \frac{f_{\mathcal{H},N+1}(\boldsymbol{x})}{\sqrt{\sum_{i=1}^{N+1}|f_{\mathcal{H},i}(\boldsymbol{x})|^2}} \boldsymbol{e}_L \tag{100}$$

$$\|\mathbb{v}(\boldsymbol{x})\| = \sqrt{\sum_{i=1}^{N+1}|f_{\mathcal{H},i}(\boldsymbol{x})|^2} \tag{101}$$

In both cases we may write

$$f_\mathcal{H}(\boldsymbol{x}) = \hat{f}_\mathcal{H}(\boldsymbol{x}) \|f_\mathcal{H}(\boldsymbol{x})\| \tag{102}$$

where

$$\hat{f}_\mathcal{H}(\boldsymbol{x}) = \begin{cases} I_L sgn\left(f_{\mathcal{H},Im}(\boldsymbol{x})\right) \\ I_L \hat{\mathbb{v}}(\boldsymbol{x}) \end{cases} \tag{103}$$

$$\|f_\mathcal{H}(\boldsymbol{x})\| = \begin{cases} |f_{\mathcal{H},Im}(\boldsymbol{x})| \\ \|\mathbb{v}(\boldsymbol{x})\| \end{cases} \tag{104}$$

The analytic signal becomes

$$f_A(\boldsymbol{x}) = \tfrac{1}{2}f(\boldsymbol{x}) + \tfrac{1}{2}\hat{f}_\mathcal{H}(\boldsymbol{x})\|f_\mathcal{H}(\boldsymbol{x})\| \tag{105}$$

We call such a signal a <u>generic analytic signal</u>. It is a refinement of the generalized analytic signal by imposing the symmetry conditions given above – all generic analytic signals are generalized analytic signals, but not all generalized analytic signals are generic analytic signals.

For a generic analytic signal we have

$$\hat{f}_{\mathcal{H}}(x)^2 = -1 \tag{106}$$

Thus the function $f_A(x)$ now looks algebraically like a complex function:

$$f_A(x) = \frac{1}{2} R(x) e^{\hat{f}_{\mathcal{H}}(x)\theta(x)} \tag{107}$$

$$R(x) = \sqrt{f^2(x) + \|f_{\mathcal{H}}(x)\|^2} \tag{108}$$

$$\theta(x) = \tan^{-1}\left(\frac{\|f_{\mathcal{H}}(x)\|}{f(x)}\right) \tag{109}$$

The unit vector $\hat{f}_{\mathcal{H}}(x)$ is called the <u>orientation vector</u> and the weighted vector $\theta(x)\hat{f}_{\mathcal{H}}(x)$ is called the <u>phase vector</u>.

Compare the generic analytic signal to the analytic signal in one dimension:

$$\begin{aligned} f_A(x) &= \frac{1}{2} f(x) + \frac{1}{2} i H(f)(x) \\ &= \frac{1}{2} f(x) + \frac{1}{2} \hat{f}_{\mathcal{H}_1}(x) |H(f)(x)| \end{aligned} \tag{110}$$

Here the function $\hat{f}_{\mathcal{H}_1}(x) = i\, sgn(H(f)(x))$ satisfies

$$\hat{f}^2_{\mathcal{H}_1}(x) = -1 \tag{111}$$

The role of the magnitude $\|f_{\mathcal{H}}(x)\|$ for a multi-dimensional generic analytic signal is analogous to the role of the magnitude $|H(f)(x)|$ of the Hilbert transform for a one-dimensional analytic signal. Similarly, the role of $\hat{f}_{\mathcal{H}}(x)$ is analogous to the role of $\hat{f}_{\mathcal{H}_1}(x) = i\, sgn(H(f)(x))$ in one dimension. For a generic analytic signal the original function can be fully recovered from $f_{\mathcal{H},Im}(x)$ or $\mathbb{v}(x)$. Consequently in some applications it may be beneficial to use $f_{\mathcal{H},Im}(x)$ or $\mathbb{v}(x)$ in place of the original function $f(x)$.

*Ordinary Analytic Signal*

In one dimension the Hilbert transform has the following property:

$$f(x) = \cos(\omega_c x) \implies f_{\mathcal{H}_1}(x) = i \sin(\omega_c x) \tag{112}$$

Accordingly we examine an analogous result in multiple dimensions. Consider

$$f(x) = \cos(\boldsymbol{\omega}_c \cdot \boldsymbol{x}) \tag{113}$$

We find

$$f_{\mathcal{H}}(\boldsymbol{x}) = \int_{\mathbb{R}^N} a_{\mathcal{H}}(\boldsymbol{\omega}) \mathcal{F}[f](\boldsymbol{\omega}) e^{I_L \boldsymbol{x}\cdot\boldsymbol{\omega}} d\boldsymbol{\omega} = \frac{a_{\mathcal{H}}(\boldsymbol{\omega}_c) e^{I_L \boldsymbol{x}\cdot\boldsymbol{\omega}} + a_{\mathcal{H}}(-\boldsymbol{\omega}_c) e^{-I_L \boldsymbol{x}\cdot\boldsymbol{\omega}}}{2} \tag{114}$$

Define the components of multiplier having even and odd symmetry with respect to the argument $\boldsymbol{\omega}$ as

$$a_{\mathcal{H},e}(\boldsymbol{\omega}) = \frac{a_{\mathcal{H}}(\boldsymbol{\omega})+a_{\mathcal{H}}(-\boldsymbol{\omega})}{2} \tag{115}$$

$$a_{\mathcal{H},o}(\boldsymbol{\omega}) = \frac{a_{\mathcal{H}}(\boldsymbol{\omega})-a_{\mathcal{H}}(-\boldsymbol{\omega})}{2} \tag{116}$$

We now have

$$f_{\mathcal{H}}(x) = a_{\mathcal{H},e}(\boldsymbol{\omega}_c) \cos(\boldsymbol{\omega}_c \cdot \boldsymbol{x}) + I_L a_{\mathcal{H},o}(\boldsymbol{\omega}_c) \sin(\boldsymbol{\omega}_c \cdot \boldsymbol{x}) \tag{117}$$

For a generic analytic signal with the idempotent described above we now find:

$$f_{\mathcal{H}}(x) = \begin{cases} I_3 m_o(\boldsymbol{\omega}_c) \sin(\boldsymbol{\omega}_c \cdot \boldsymbol{x}) \\ P_e(\boldsymbol{\omega}_c) \cos(\boldsymbol{\omega}_c \cdot \boldsymbol{x}) I_L \boldsymbol{e}_L + I_L \boldsymbol{v}_o(\boldsymbol{\omega}_c) \sin(\boldsymbol{\omega}_c \cdot \boldsymbol{x}) \end{cases} \tag{118}$$

When the idempotent is generated by a scalar function, the generic analytic function whose multiplier has odd symmetry with respect to its argument already has the desired behavior analogous to (112). However, when the idempotent is generated by vector and pseudovector functions, to obtain the desired behavior we must eliminate the even-grade component of the multiplier by setting $P_e(\boldsymbol{\omega}) \equiv 0$. This in turn implies the odd-grade component of the multiplier is a unit vector with odd symmetry. We then find

$$f_{\mathcal{H}}(\boldsymbol{x}) = I_L \hat{\boldsymbol{v}}(\boldsymbol{\omega}_c) \sin(\boldsymbol{\omega}_c \cdot \boldsymbol{x}) \tag{119}$$

We now define an <u>ordinary analytic signal</u> to be any generic analytic signal restricted in this way. Algebraically it continues to look like a complex function. When the idempotent is generated by a unit vector the component $f_{\mathcal{H},N+1}(\boldsymbol{x})$ vanishes and the extended Hilbert transform becomes

$$I_L \mathbb{v}(\boldsymbol{x}) = I_L \sum_{i=1}^N f_{\mathcal{H},i}(\boldsymbol{x}) \, \boldsymbol{e}_i \tag{120}$$

$$\|\mathbb{v}(\boldsymbol{x})\| = \sqrt{\sum_{i=1}^N |f_{\mathcal{H},i}(\boldsymbol{x})|^2} \tag{121}$$

$$\hat{\mathbb{v}}(\boldsymbol{x}) = \sum_{i=1}^N \frac{f_{\mathcal{H},i}(\boldsymbol{x})}{\sqrt{\sum_{i=1}^N |f_{\mathcal{H},i}(\boldsymbol{x})|^2}} \, \boldsymbol{e}_i \tag{122}$$

We can recover the original signal from $f_{\mathcal{H},Im}(\boldsymbol{x})$ or $\|\mathbb{v}(\boldsymbol{x})\|, \hat{\mathbb{v}}(\boldsymbol{x})$. Accordingly one may characterize an ordinary analytic signal either by a magnitude and angle given by $R(\boldsymbol{x})$ and $\theta(\boldsymbol{x})$ as defined previously or – when the idempotent is generated by a unit vector – by a magnitude and unit vector given by $\|\mathbb{v}(\boldsymbol{x})\|$ and $\hat{\mathbb{v}}(\boldsymbol{x})$. We explore this idea further below.

The framework set forth herein defines a generalized analytic signal in terms of an idempotent. It then imposes symmetry conditions on the Fourier multiplier associated with the idempotent to define a subset of signals called generic analytic signals. It then imposes a further condition on the idempotent of a generic analytic signal to define a subset of signals called ordinary analytic signals. The hierarchy is

Generalized Analytic Signals ⊃ Generic Analytic Signals ⊃ Ordinary Analytic Signals

**Examples of Two-Dimensional Analytic Signals**

We now let $N = 2, L = 3$. In this case we have

$$I_L = I_3 = e_1 e_2 e_3 \tag{123}$$

$$I_{L-1} = I_2 = e_1 e_2 = I_3 e_3 \tag{124}$$

We begin by examining an idempotent formed from a scalar function $m(\boldsymbol{\omega})$. In particular let $m(\boldsymbol{\omega})$ be a Fourier multiplier given by

$$m(\boldsymbol{\omega}) = \frac{\frac{\omega_1}{|\omega_1|} + \frac{\omega_2}{|\omega_2|} + \frac{\omega_1 \omega_2}{|\omega_1||\omega_2|} - 1}{2} \tag{125}$$

It is straightforward to show $m^2(\boldsymbol{\omega}) = 1$. The associated idempotent is given by

$$\psi_{\mathcal{H}}(\boldsymbol{\omega}) = \left(\frac{1 + \frac{\omega_1}{|\omega_1|}}{2}\right)\left(\frac{1 + \frac{\omega_2}{|\omega_2|}}{2}\right) \tag{126}$$

Hahn calls this function a two-dimensional unit step function [7]. In the frequency domain the resulting generalized analytic signal is given by

$$\mathcal{F}[f_A](\boldsymbol{\omega}) = \left(\frac{1 + \frac{\omega_1}{|\omega_1|}}{2}\right)\left(\frac{1 + \frac{\omega_2}{|\omega_2|}}{2}\right) \mathcal{F}[f](\boldsymbol{\omega}) \tag{127}$$

The only non-vanishing frequency components are those in the first quadrant $\omega_1, \omega_2 \geq 0$. Define the following functions:

$$f_{H1}(\boldsymbol{x}) = \mathcal{F}^{-1}\left[\frac{\omega_1}{|\omega_1|} \mathcal{F}[f](\boldsymbol{\omega})\right](\boldsymbol{x}) \tag{128}$$

$$f_{H2}(\boldsymbol{x}) = \mathcal{F}^{-1}\left[\frac{\omega_2}{|\omega_2|} \mathcal{F}[f](\boldsymbol{\omega})\right](\boldsymbol{x}) \tag{129}$$

$$f_{HT}(\boldsymbol{x}) = \mathcal{F}^{-1}\left[\frac{\omega_1}{|\omega_1|}\frac{\omega_2}{|\omega_2|} \mathcal{F}[f](\boldsymbol{\omega})\right](\boldsymbol{x}) \tag{130}$$

These functions are the <u>partial and total Hilbert transforms</u> of $f$ and are real functions of $\boldsymbol{x}$. In the spatial domain the generalized analytic signal becomes:

$$f_A(\boldsymbol{x}) = \frac{1}{4}\left(f(\boldsymbol{x}) + f_{HT}(\boldsymbol{x})\right) + I_3 \frac{1}{4}\left(f_{H1}(\boldsymbol{x}) + f_{H2}(\boldsymbol{x})\right) \tag{131}$$

Except for scaling, this signal is the two-dimensional single-orthant signal of Hahn [7].[5] It is a <u>generalized</u> analytic signal.

---

[5] Hahn's original formulation is in terms of complex numbers. Here the unit pseudoscalar plays the role of the unit imaginary.

Now consider idempotents formed from vector and bivector functions. For example let a multiplier be given by

$$v(\omega) = \frac{\omega_1}{|\omega_1|} e_1 + \frac{\omega_2}{|\omega_2|} e_2 \tag{132}$$

$$P(\omega) = s(\omega) = \frac{\omega_1}{|\omega_1|} \frac{\omega_2}{|\omega_2|} \tag{133}$$

The odd-grade component $v(\omega)$ has odd symmetry with respect to its argument whereas the even-grade component $P(\omega)e_1e_2$ has even symmetry. This gives rise to a <u>generic</u> analytic signal having idempotent

$$\psi_\mathcal{H}(\omega) = \left( \frac{1 + \frac{\omega_1}{|\omega_1|} e_1 + \frac{\omega_2}{|\omega_2|} e_2 + \frac{\omega_1}{|\omega_1|} \frac{\omega_2}{|\omega_2|} e_1 e_2}{2} \right) = 2 \left( \frac{1 + \frac{\omega_1}{|\omega_1|} e_1}{2} \right) \left( \frac{1 + \frac{\omega_2}{|\omega_2|} e_2}{2} \right) \tag{134}$$

In the spatial domain the analytic signal becomes

$$f_A(x) = \frac{1}{2} f(x) + \frac{1}{2} f_{H1}(x) I_3 e_1 + \frac{1}{2} f_{H2}(x) I_3 e_2 + \frac{1}{2} f_{HT}(x) e_1 e_2 \tag{135}$$

This generic analytic signal $f_A$ has the same components as the two-dimensional hypercomplex signal of Bulow et al. [1, eq. 46], who express it in the algebra of quaternions. From this presentation it is evident that the two-dimensional hypercomplex signal and the Hahn signal are related [8]. However, whereas in our hierarchy of analytic signals the Hahn signal is a generalized signal, the hypercomplex signal is a member of the smaller subset of generic signals.

For a two-dimensional generic analytic signal we find in general

$$\hat{v}(x) = \frac{f_{\mathcal{H},1}(x)}{\|v(x)\|} e_1 + \frac{f_{\mathcal{H},2}(x)}{\|v(x)\|} e_2 + \frac{f_{\mathcal{H},3}(x)}{\|v(x)\|} e_3 \tag{136}$$

Define the following angles:

$$\sigma(x) = \tan^{-1} \left( \frac{f_{\mathcal{H},2}(x)}{f_{\mathcal{H},1}(x)} \right) \tag{137}$$

$$\kappa(x) = \sin^{-1} \left( \frac{f_{\mathcal{H},3}(x)}{\|v(x)\|} \right) \tag{138}$$

With these definitions we now have

$$\hat{v}(x) = \cos \kappa(x) \cos \sigma(x) e_1 + \cos \kappa(x) \sin \sigma(x) e_2 + \sin \kappa(x) e_3 \tag{139}$$

The angle $\sigma(x)$ is called the <u>orientation angle</u> and the angle $\kappa(x)$ is called the <u>elevation angle</u>. They are illustrated in Figure 1.

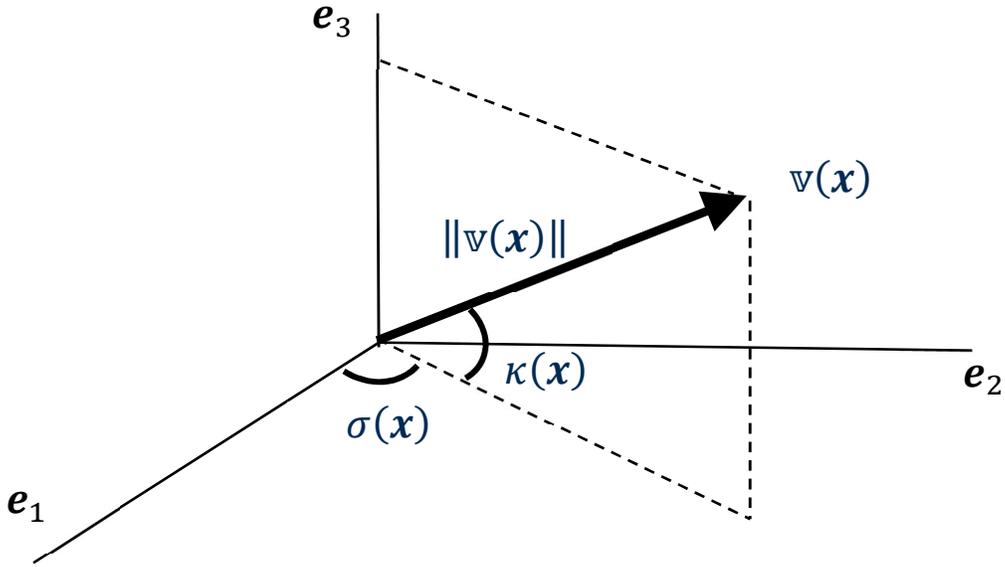

**Figure 1.** Orientation and Elevation Angles of Extended Hilbert Transform $\mathbb{v}(x)$

The extended Hilbert transform associated with a two-dimensional generic analytic signal has this geometric interpretation. When the generic signal is restricted to an ordinary analytic signal, the elevation angle $\kappa(x)$ vanishes.

To obtain a new result, we scale the vector component of the hypercomplex signal multiplier:

$$v(\omega) = \frac{sgn(\omega_1)e_1 + sgn(\omega_2)e_2}{\sqrt{2}} \qquad (140)$$

This scaling makes $v(\omega)$ a unit vector and eliminates the even-grade component. We thus obtain an <u>ordinary</u> analytic signal given by

$$f_A(x) = \tfrac{1}{2}f(x) + \tfrac{1}{2\sqrt{2}}f_{H1}(x)e_1 + \tfrac{1}{2\sqrt{2}}f_{H2}(x)e_2 \qquad (141)$$

This signal may reasonably be called a modified hypercomplex signal. Geometrically one can think of it obtaining it from the hypercomplex signal by setting the elevation angle $\kappa(x)$ to zero and rescaling so that $\|\mathbb{v}(x)\| = 1$.

Now examine a Fourier multiplier with

$$v(\boldsymbol{\omega}) = \frac{\boldsymbol{\omega}}{|\boldsymbol{\omega}|} = \hat{\boldsymbol{\omega}} \tag{142}$$

$$P(\boldsymbol{\omega}) = 0 \tag{143}$$

It is an odd function of $\boldsymbol{\omega}$ and hence gives rise to a <u>ordinary</u> analytic signal:

$$\mathcal{F}[f_A] = \left(\frac{1+\hat{\boldsymbol{\omega}}}{2}\right)\mathcal{F}[f] \tag{144}$$

Define

$$f_{\mathcal{R}1}(\boldsymbol{x}) = \mathcal{F}^{-1}\left[\frac{\omega_1}{|\boldsymbol{\omega}|}\mathcal{F}[f](\boldsymbol{\omega})\right](\boldsymbol{x}) \tag{145}$$

$$f_{\mathcal{R}2}(\boldsymbol{x}) = \mathcal{F}^{-1}\left[\frac{\omega_2}{|\boldsymbol{\omega}|}\mathcal{F}[f](\boldsymbol{\omega})\right](\boldsymbol{x}) \tag{146}$$

The functions $f_{\mathcal{R}1}, f_{\mathcal{R}2}$ are the <u>Riesz transforms</u> of $f$. They are real functions of $\boldsymbol{x}$. In the spatial domain the ordinary analytic signal becomes

$$f_A(\boldsymbol{x}) = \tfrac{1}{2}f(\boldsymbol{x}) + \tfrac{1}{2}f_{\mathcal{R}1}(\boldsymbol{x})\, I_3 \boldsymbol{e}_1 + \tfrac{1}{2}f_{\mathcal{R}2}(\boldsymbol{x})\, I_3 \boldsymbol{e}_2 \tag{147}$$

This ordinary analytic signal $f_A$ has the same components as the monogenic signal of Felsberg et al. [3, eq. 13], who express it in the algebra of quaternions.

In addition to encompassing the two-dimensional Hahn, monogenic, hypercomplex, and modified hypercomplex analytic signals, the two-dimensional generalized analytic signal defined above gives a mechanism for forming other generalized analytic signals through the choice of $v(\boldsymbol{\omega})$. For example, consider a general vector $\boldsymbol{v}_G$:

$$\boldsymbol{v}_G(\omega_1, \omega_2) = A\frac{v_1(\omega_1,\omega_2)}{\sqrt{v_1^2(\omega_1,\omega_2)+v_2^2(\omega_1,\omega_2)}}\boldsymbol{e}_1 + A\frac{v_2(\omega_1,\omega_2)}{\sqrt{v_1^2(\omega_1,\omega_2)+v_2^2(\omega_1,\omega_2)}}\boldsymbol{e}_2 \tag{148}$$

Here $v_1(\omega_1, \omega_2), v_2(\omega_1, \omega_2)$ are functions to be specified and $A = |\boldsymbol{v}_G(\omega_1, \omega_2)| > 0$. The Fourier multiplier becomes:

$$a_{\mathcal{H}}(\omega_1, \omega_2) = A\frac{v_1(\omega_1,\omega_2)}{\sqrt{v_1^2(\omega_1,\omega_2)+v_2^2(\omega_1,\omega_2)}}\boldsymbol{e}_1 + A\frac{v_2(\omega_1,\omega_2)}{\sqrt{v_1^2(\omega_1,\omega_2)+v_2^2(\omega_1,\omega_2)}}\boldsymbol{e}_2 + s(\omega_1,\omega_2)\sqrt{A^2-1}\,\boldsymbol{e}_1\boldsymbol{e}_2 \tag{149}$$

We obtain the previously-discussed signals with the following choices:

**Table 1. Parameter Choices for Different Signals**

| | |
|---|---|
| Monogenic | $A = 1$ |
| | $v_1(\omega_1, \omega_2) = \omega_1$ |
| | $v_2(\omega_1, \omega_2) = \omega_2$ |
| Modified Hypercomplex | $A = 1$ |
| | $v_1(\omega_1, \omega_2) = sgn(\omega_1)$ |
| | $v_2(\omega_1, \omega_2) = sgn(\omega_2)$ |
| Hypercomplex | $A = \sqrt{2}$ |
| | $v_1(\omega_1, \omega_2) = sgn(\omega_1)$ |
| | $v_2(\omega_1, \omega_2) = sgn(\omega_2)$ |
| | $s(\omega_1, \omega_2) = sgn(\omega_1)sgn(\omega_2)$ |

For examples of other multipliers, consider the following implementation of this particular model:

$$A = 1 \tag{150}$$

$$v_1(\omega_1, \omega_2) = A_1 sgn(\omega_1) \left(\frac{|\omega_1|}{|\boldsymbol{\omega}|}\right)^{\alpha_1} + B_1 sgn(\omega_2) \left(\frac{|\omega_2|}{|\boldsymbol{\omega}|}\right)^{\beta_1} \tag{151}$$

$$v_2(\omega_1, \omega_2) = A_2 sgn(\omega_1) \left(\frac{|\omega_1|}{|\boldsymbol{\omega}|}\right)^{\alpha_2} + B_2 sgn(\omega_2) \left(\frac{|\omega_2|}{|\boldsymbol{\omega}|}\right)^{\beta_2} \tag{152}$$

Because $v_1(\omega_1, \omega_2), v_2(\omega_1, \omega_2)$ are odd the resulting signal will be an ordinary analytic signal. The choice $\alpha_1 = \beta_2 = 1, A_1 = B_2 = 1, A_2 = B_1 = 0$ reproduces the monogenic signal and $\alpha_1 = \beta_2 = 0, A_1 = B_2 = 1, A_2 = B_1 = 0$ reproduces the modified hypercomplex signal. If we think of the multiplier $\boldsymbol{v}_G(\boldsymbol{\omega})$ as a vector field, then we have the following plots of this vector field for various choices of the parameters $A_1, A_2, B_1, B_2, \alpha_1, \alpha_2, \beta_1, \beta_2$. It is immediately evident that $\boldsymbol{v}_G(-\boldsymbol{\omega}) = -\boldsymbol{v}_G(\boldsymbol{\omega})$ in each case.

Monogenic Signal

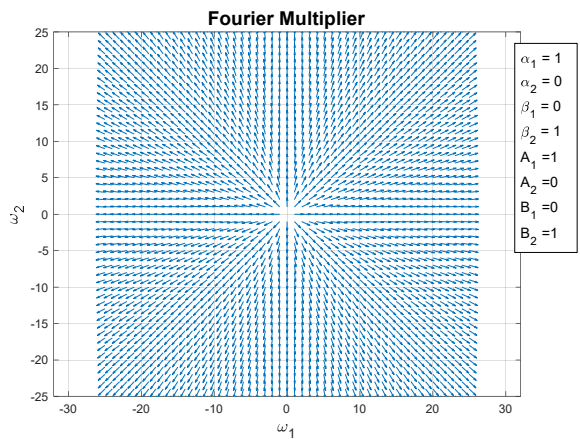

Perpendicular Monogenic Signal

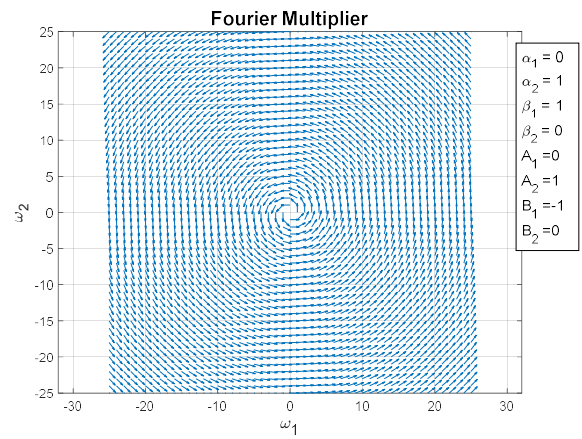

Modified Hypercomplex Signal

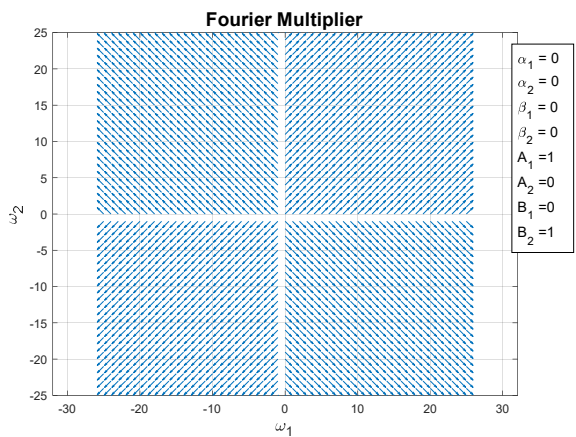

Perpendicular Modified Hypercomplex Signal

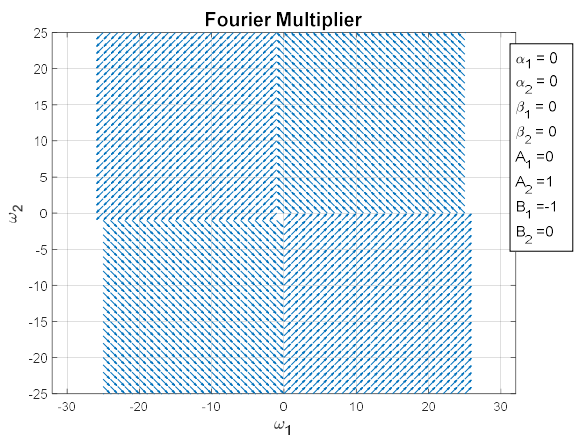

**Figure 2.** Vector Fields Associated with Fourier Multipliers of Monogenic and Modified Hypercomplex Signals

Other models can be defined.

**Example Application to Images**

In this section we apply the previous ideas to two-dimensional signals in the form of images. Consider the following image and the vector field associated with the extended Hilbert transform (i.e. $-I_3 f_{\mathcal{H}}(x) = \mathbb{v}(x)$) obtained from the monogenic signal:

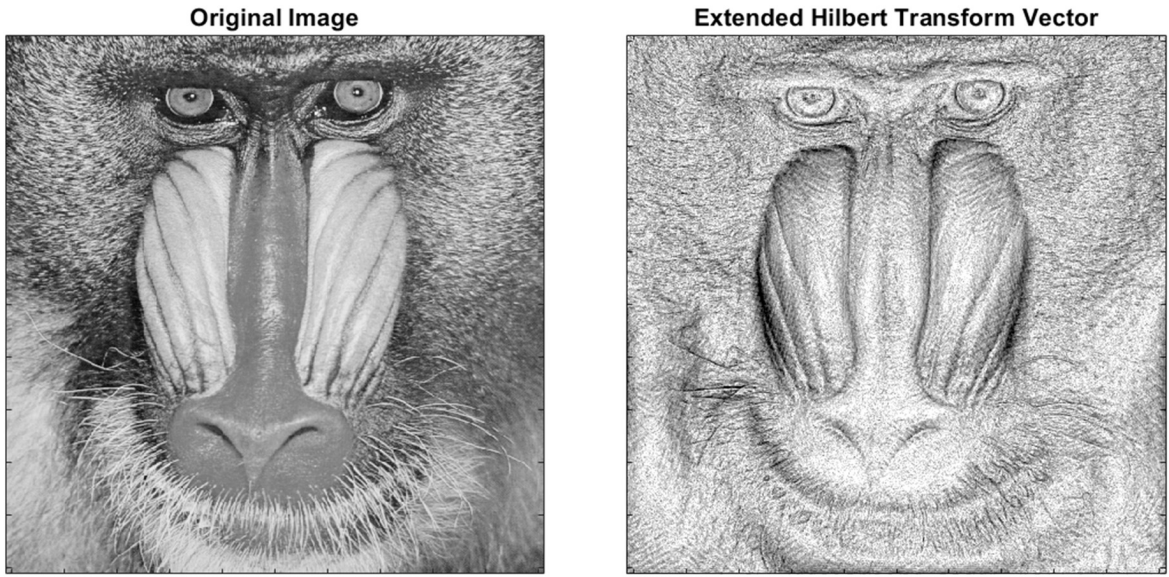

**Figure 3.** Monogenic Signal – Original Image and Vector Field $\mathbb{v}(x)$ of Extended Hilbert Transform

The ordinary analytic signal comprises the original image and the vector field of the extended Hilbert transform. The magnitude and phase vector field of the analytic signal are also shown in Figure 4:

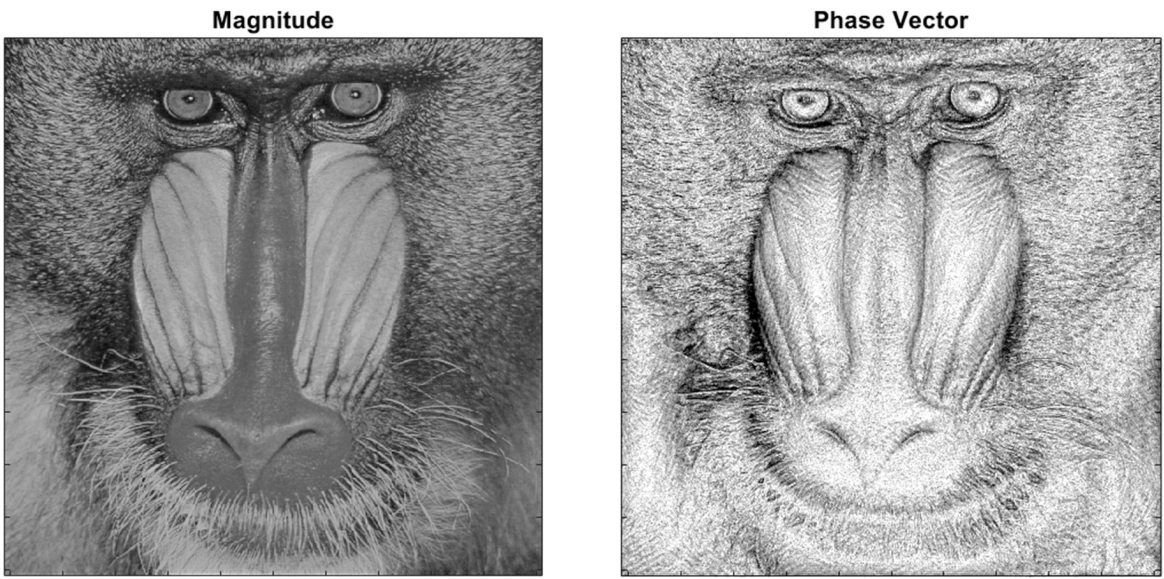

**Figure 4.** Monogenic Signal – Magnitude $R(x)$ and Phase Vector $\theta(x)\hat{\mathbb{v}}(x)$ Corresponding to Figure 3.

As emphasized above, the vector field of the extended Hilbert transform contains all of the information in the original figure and can perhaps be used in lieu of the original figure in some applications. It may be characterized in terms of two components in the $e_1$ and $e_2$ directions:

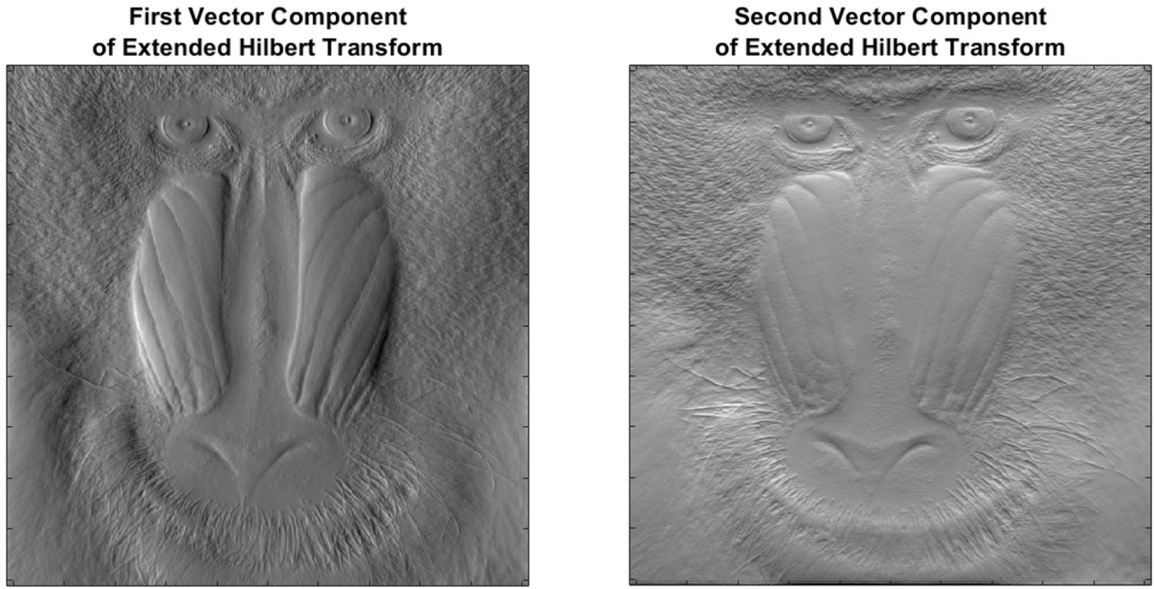

**Figure 5.** Components Associated with the Extended Hilbert Transform – $f_{\mathcal{H},1}(x), f_{\mathcal{H},2}(x)$

Alternatively the vector field defining the extended Hilbert transform may be characterized by a magnitude and a unit orientation vector (we could have plotted the orientation angle $\sigma(x)$ in lieu of the unit orientation vector field.).

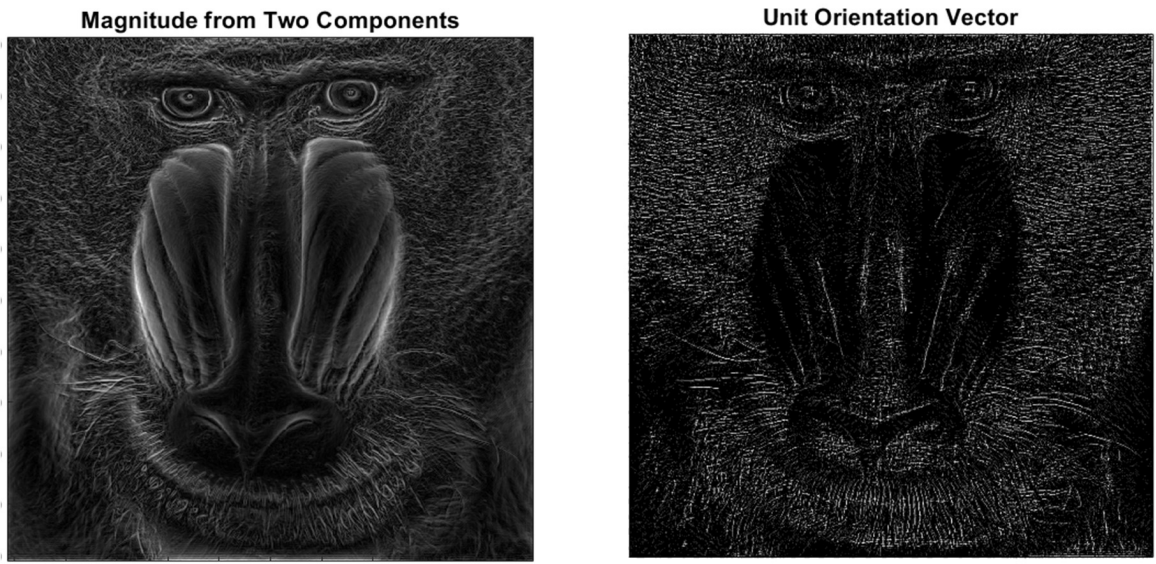

**Figure 6.** Components Associated with the Extended Hilbert Transform – $\|\mathbb{v}(x)\|, \hat{\mathbb{v}}(x)$

Each of these various formulations appears to capture essential features of the original image. Therefore it is of some interest to investigate applications in which one or more of these formulations may be used effectively in place of the original image.

As illustrated in the following diagram, the original signal $f(x)$ can be fully recovered from the extended Hilbert transform by applying the operator $\mathcal{H}$ (the vector field multiplier operates in the frequency domain, so it is understood that the components of the extended Hilbert transform are first transformed via a Fourier transform into the frequency domain and then after the multiplier is applied are transformed back into the spatial domain – this process is understood and not explicitly shown in the figure):

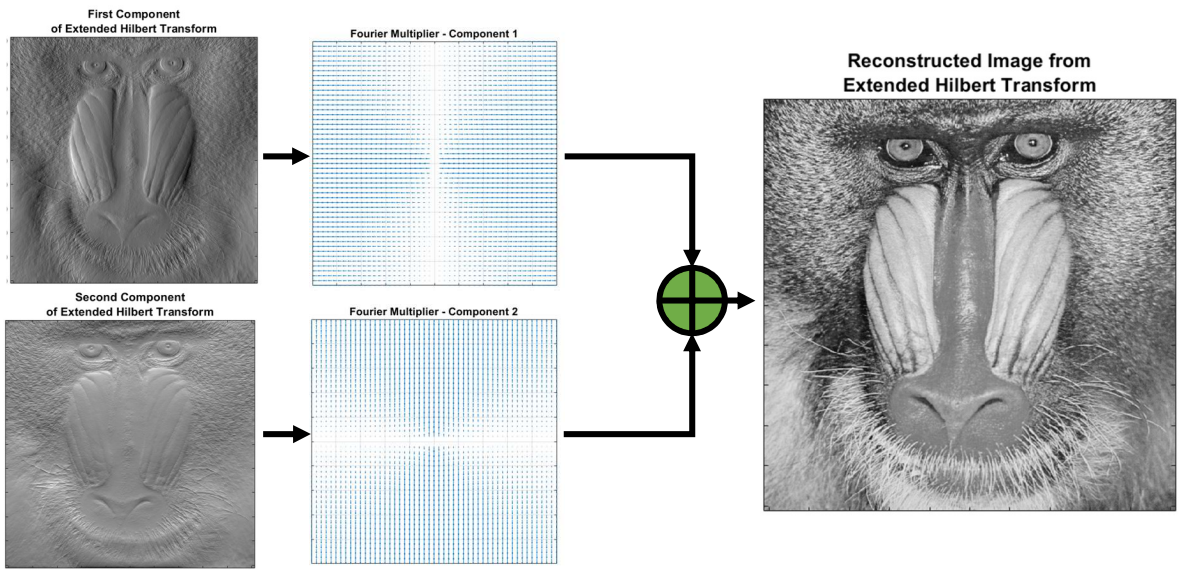

**Figure 7.** Reconstruction of Original Signal from Extended Hilbert Transform $\mathbb{v}(x)$

The following result was obtained by reconstructing an image from the unit orientation vector alone:

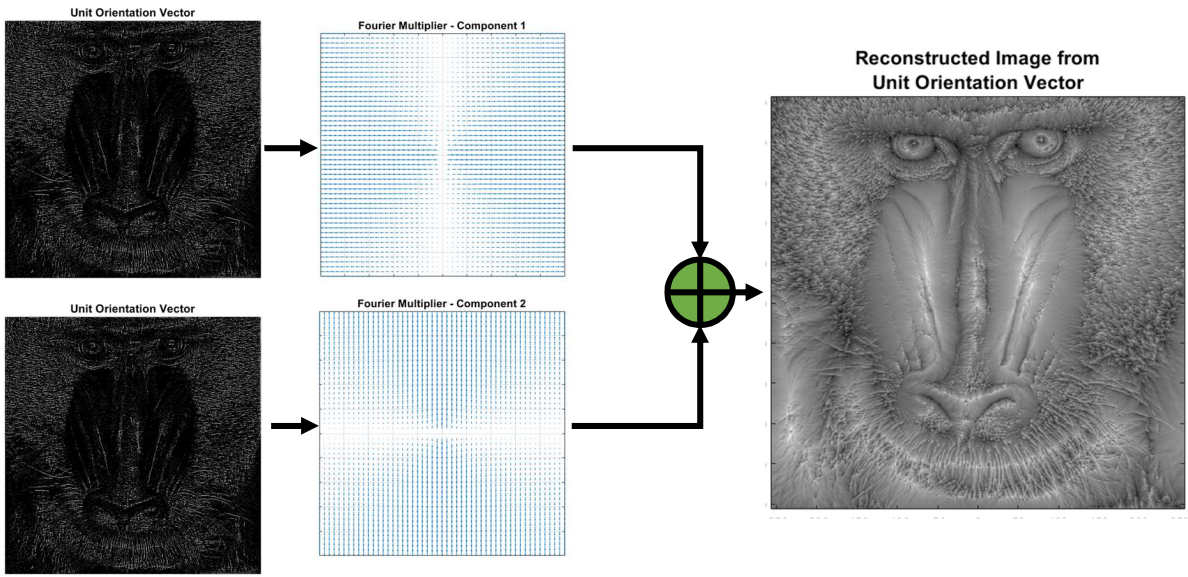

**Figure 8.** Reconstruction of Signal from Unit Orientation Vector $\hat{\mathbb{v}}(x)$

Although use of the unit orientation vector alone does not fully recover the original image, it nonetheless appears to capture significant features of the original image.

Finally, we note that other choices for the Fourier multiplier are possible. The following figure shows the components of the extended Hilbert transform resulting from use of a random vector field as the multiplier – at each point in the frequency plane a unit vector is oriented randomly except for the constraint that the symmetry condition required to insure an ordinary analytic signal is satisfied:

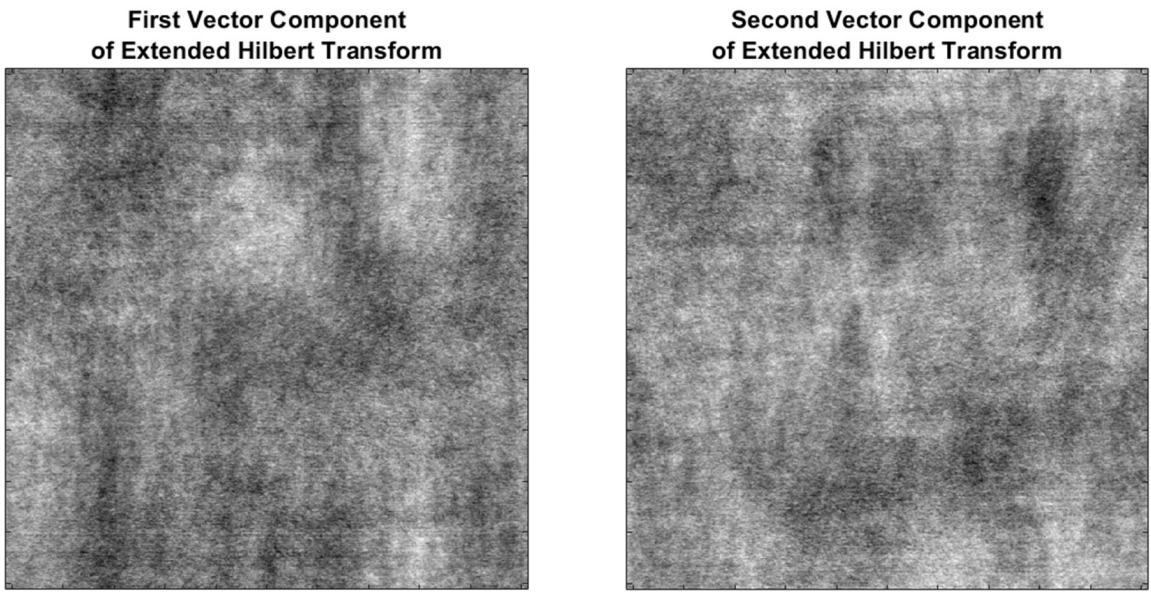

**Figure 9.** Extended Hilbert Transform with Random Idempotent

Despite the random appearance of these components, the original image is fully recovered by applying the operator $\mathcal{H}$:

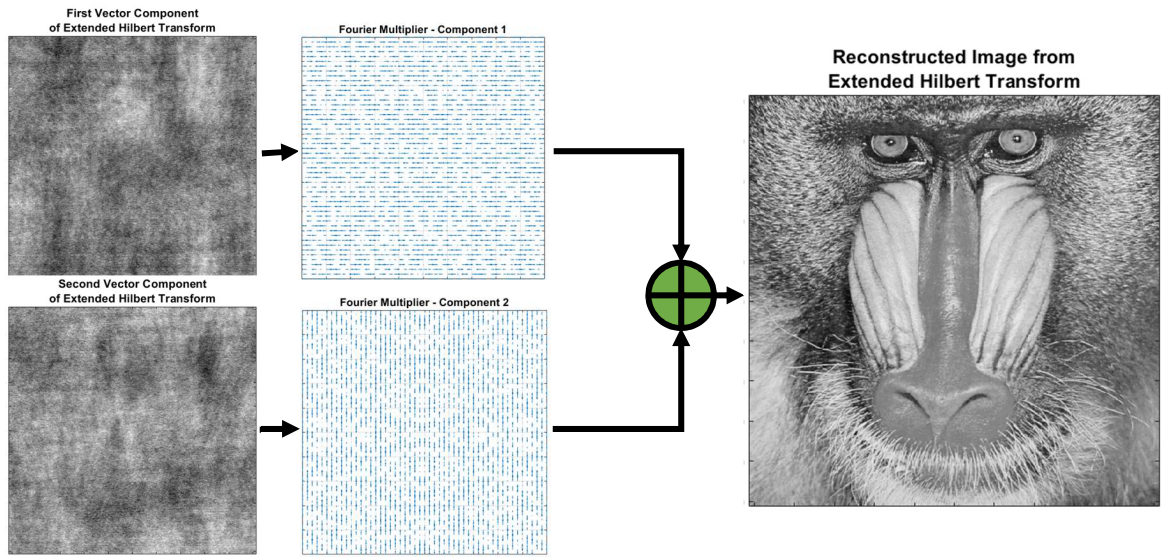

**Figure 10.** Reconstruction of Original Signal with Random Fourier Multiplier

The following figure shows the reconstruction of the image in this case using only the unit orientation vector, which is also shown along with a zoomed in portion from the center of the field to show that it indeed appears to be random:

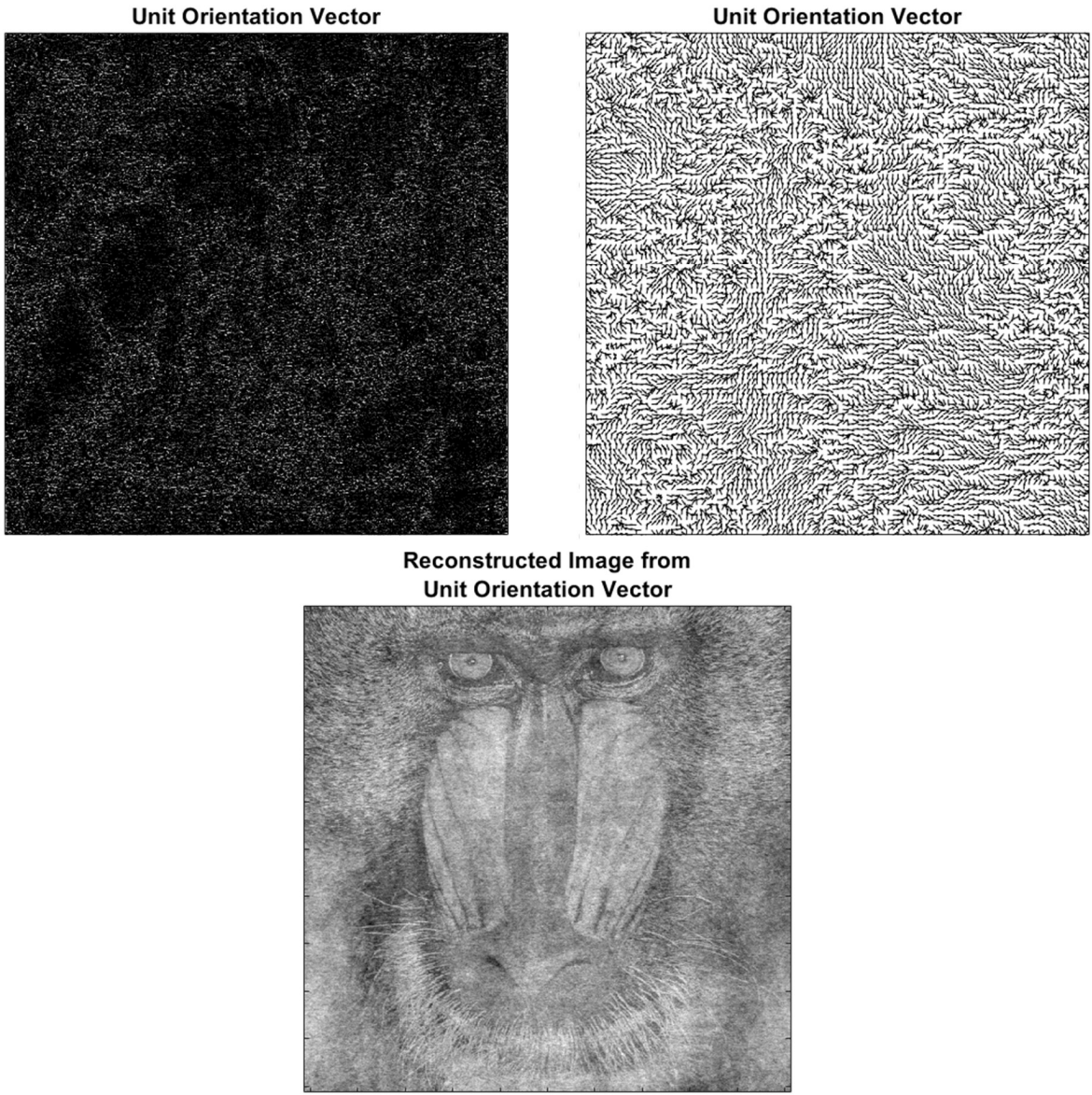

**Figure 11.** Reconstruction of Original Signal with Random Fourier Multiplier Using Only Unit Orientation Vector

Although the image is not fully recovered, the unit orientation vector nonetheless encodes significant features of the original image.

**Conclusion and Suggestions for Further Work**

Herein we have set forth a general framework for extending the one-dimensional analytic signal to multiple dimensions and have shown how this framework encompasses the Hahn single-orthant signal, the hypercomplex signal of Bulow et al., and the monogenic signal of Felsberg et al. We have also shown how it extends these previously known approaches to enable new definitions of a multidimensional analytic signal. From a mathematical perspective, several issues merit further exploration:

1. Under what conditions does the extended Hilbert operator $\mathcal{H}$ exist;
2. Under what conditions does an associated convolution kernel exist;
3. In one dimension the Hilbert transform is essentially the only singular integral [12]. Accordingly it seems natural to ask what role if any is played in this framework by multidimensional versions of singular integrals such as Calderon-Zygmund singular integral operators. In this regard the work of Brackx et al. is particularly pertinent [13-15].
4. The analytic signal in one dimension may be obtained as the boundary value of a holomorphic function in the complex plane. Similarly, the hypercomplex signal, the monogenic signal, and Hahn's single-orthant signal can each respectively be obtained from suitably defined Riemann-Hilbert boundary value problems [9, 10]. It is therefore of some interest to determine under what conditions (if any) the generalized analytic signal defined herein satisfies a suitably defined boundary-value problem.

From an engineering perspective, the question arises as to how to use the various forms of the multidimensional analytic signal set forth herein. For example, one possible application might be to encrypt images by using a random multiplier vector field to define an encryption key. Other applications of the general approach await to be identified. In addition, it is of some interest to ascertain to what extent the original signal can be recovered from, for example, selected components of the extended Hilbert transform. In the two-dimensional case we showed how the unit orientation vector field can be used to recover a version of the original signal; the question arises as to what extent can other components be used to recover useful forms of the original signal.

**References**

[1] Bulow,T., and Sommer, G., "Hypercomplex Signals – A Novel Extension of the Analytic Signal to the Multidimensional Case," *IEEE Transactions on Signal Processing*, **49**(11), Nov. 2001, 2844-2852

[2] Bulow, T., "Hypercomplex Spectral Signal Representations for the Processing and Analysis of Images," Ph.D. dissertation, Christian Albrechts University, Kiel, Germany, 1999

[3] Felsberg, M., and Sommer, G., "The Monogenic Signal," *IEEE Transactions on Signal Processing*, **49**(12), Dec. 2001, 3136-3144

[4] Felsberg, M., "Low-Level Image Processing with the Structure Multivector," Ph.D. dissertation, Christian Albrechts University, Kiel, Germany, 2002

[5] Brackx, F., Delanghe, R., and Sommen, F., CLIFFORD ANALYSIS, Boston, Pitman, 1982

[6] Brackx, F., De Knock, B., and De Schepper, H., "Hilbert Transforms in Clifford Analysis," in GEOMETRIC ALGEBRA COMPUTING, Bayro-Corrochano, E., and Scheuermann, G., eds., Springer-Verlag, London, 163-187, 2010

[7] Hahn, S.L., "Multidimensional Complex Signals with Single-Orthant Spectra," *Proceedings of the IEEE*, **80**(8), Aug. 1992, 1287-1300

[8] Hahn, S.L., and Snopek, K.M., "Comparison of Properties of Analytic, Quaternionic and Monogenic 2-D Signals", *World Scientific and Engineering Academy and Society Transactions on Computers*, **3**(3), January 2004

[9] Cerejeiras, P., Kahler, U, "Monogenic Signal Theory," in OPERATOR THEORY, Alpay, D., ed., Springer Basel, 2015, 1701-1724

[10] Hahn, S.L., "Complex Signals with Single-Orthant Spectra as Boundary Distributions of Multidimensional Analytic Functions," Bulleting of the Polish Academy of Science, Technical Sciences, **2**, 2003

[11] Kothe, U, and Felberg, M., "Riesz-Transforms vs Derivatives: On the Relationship between the Boundary Tensor and the Energy Tensor," International Conference on Scale-Space Theories in Computer Vision, 2005, 179-191

[12] Krantz, S.G., EXPLORATIONS IN HARMONIC ANALYSIS, Birkhauser, 2009, p. 54

[13] Brackx, F., and De Schepper, H., "On the Fourier Transform of Distributions and Differential Operators in Clifford Analysis," *Complex Variables*, **49**(15), December 2004, 1079-91

[14] Brackx, F., and De Schepper, H., "Convolution Kernels in Clifford Analysis: Old and New," *Mathematical Methods in the Applied Sciences*, **28**, July 2005, 2173-82

[15] Brackx, F., De Knock, B., and De Schepper, H., "On the Fourier Spectra of Distributions in Clifford Analysis," *Chinese Annals of Mathematics, Series B*, **27B**(5), 2006, 495-506